\documentclass[twocolumn]{webofc}
\usepackage[varg]{txfonts}   
\usepackage{graphicx}
\usepackage{changepage}
\usepackage[symbol]{footmisc}
\usepackage{amsmath} 
\usepackage{amssymb} 
\usepackage{slashed}
\usepackage{inputenc}
\usepackage{color}

\def\bea{\begin{eqnarray}}
\def\eea{\end{eqnarray}}
\def\bean{\begin{equation*}}
\def\eean{\end{equation*}}

\begin{document}
\title{Dark Side of the Neutron?\thanks{{Plenary talk presented at the International Workshop on Particle Physics at 
Neutron Sources 2018, Grenoble, France, May 24-26, 2018; based on: B. Fornal and B. Grinstein, Phys.\,Rev.\,Lett.\,120, 191801 (2018) \cite{Fornal:2018eol}; speaker: B. Fornal. \vspace{1mm} }}}
%
%

\author{\firstname{Bartosz} \lastname{Fornal}\inst{1}\fnsep\thanks{\email{bfornal@ucsd.edu}}  \and
        \firstname{Benjam\'{i}n} \lastname{Grinstein}\inst{1}\fnsep\thanks{\email{bgrinstein@ucsd.edu}}}

\institute{\!\!Department of Physics, 
 University of California, San Diego, 
  9500 Gilman Drive, La Jolla, CA 92093, USA}

\abstract{We discuss our recently proposed interpretation of the discrepancy between the bottle and beam neutron lifetime experiments as a sign of a dark sector. The difference between the outcomes of the two types of measurements is explained by the existence of a neutron dark decay channel with a branching fraction 1\%. Phenomenologically consistent particle physics models for the neutron dark decay can be constructed and they involve a strongly self-interacting dark sector. We elaborate on the theoretical developments around this idea and describe the efforts undertaken to verify it experimentally.}
\maketitle
\section{Neutron Lifetime Discrepancy}
Although the neutron has been known for almost a century, the latest experimental results suggest that it may still be hiding a deep secret. In the currently established framework of particle physics, the Standard Model, the neutron decays almost exclusively  through beta decays, involving
\bea
n \rightarrow p + e^- \!+ \bar\nu_e 
\eea
and radiative corrections to this process. A calculation of the neutron lifetime in the Standard Model yields  \cite{Marciano:2005eca}
\bea\label{life3}
\tau_n^{\rm SM} = \frac{4908.7(1.9)\,{\rm s}}{|V_{ud}|^2(1+3g_A^2)} \ ,
\eea
where $g_A$ is the axial-vector coefficient in beta decay,  i.e., $\mathcal{M} = \tfrac{1}{\sqrt2}\,{G_F V_{ud}\, g_V}\big[\bar{p} \,\gamma_\mu n - g_A \bar{p}\, \gamma_5\gamma_\mu n\big] \left[  \bar{e} \,\gamma^\mu (1-\gamma_5) \nu \right]$. 
By using the average values of $V_{ud}$ and $g_A$ extracted from experiments and adopted by the Particle Data Group (PDG) \cite{Tanabashi:2018oca}, one arrives at the neutron lifetime in the range $875.3 \ {\rm s}< \tau_n < 891.2 \ {\rm s}$ within $3\,\sigma$. In turn, a recent lattice QCD calculation of $g_A$ \cite{Chang:2018uxx,Berkowitz:2018gqe} gave $\tau_n = 885 \pm 15 \ {\rm s}$.

There are two qualitatively  different approaches to measuring the neutron lifetime: the bottle experiments and the beam experiments.

The bottle method relies on trapping neutrons in a container and counting them at several points in time. The decaying exponential
\bea
N_n(t) = N_n(0) \,\exp\left({-{t}/{\tau_n}}\right)
\eea
is then fit to the data points $N_n(t)$, and  $\tau_n^{\rm bottle}$ is read off. Such measurement yields  the total neutron lifetime and is  independent of the actual decay channels. 
The average bottle result quoted by the PDG  and based on five  experiments \cite{Mampe,Serebrov:2004zf,Pichlmaier:2010zz,Steyerl:2012zz,Arzumanov:2015tea}  is
\bea
\tau_n^{\rm bottle}  = 879.6 \pm 0.6  \ {\rm s} \ . 
\eea
The two most recent bottle experiments \cite{Serebrov:2017bzo,Pattie:2017vsj} provided values for $\tau_n$ within $2\, \sigma$ of this average.

A different approach has been implemented in beam experiments, where the neutron lifetime is determined by counting the protons ($N_p$) resulting from neutron decays. Estimating also the number of neutrons in the beam ($N_n$) that  those protons originate from, $\tau_n^{\rm beam}$ is given by
\bea\label{onee3}
{\tau^{\rm beam}_n} = -\frac{N_n}{{d N_p}/{dt}} = \frac{\tau_n}{{\rm Br}(n\rightarrow p + {\rm anything})} \ .
\eea
In the Standard Model   $\,{\rm Br}(n\rightarrow p \,+\, {\rm anything}) = 100\%$, implying the two lifetimes are the same, $\tau^{\rm beam}_n= \tau^{\rm bottle}_n$. This equality no longer hold if other, beyond Standard Model neutron  decay channels  not involving a proton in the final state are allowed. In such  a case the branching fraction 
 ${\rm Br}(n\rightarrow p + {\rm anything}) < 100\%$ and, given Eq.\,(\ref{onee3}),
\bea\label{ineq3}
\tau^{\rm beam}_n   > \tau^{\rm bottle}_n \ . 
\eea
The average based on two beam experiments \cite{Byrne:1996zz,Yue:2013qrc} (see also Ref.\,\cite{Nico:2004ie} for the original data used in Ref.\,\cite{Yue:2013qrc}) and adopted by the PDG is
\bea
\tau_n^{\rm beam}  = 888.0 \pm 2.0  \ {\rm s} \ .
\eea
This represents a $4.0 \, \sigma$ discrepancy  with $\tau_n^{\rm bottle}$ and hints that the inequality in Eq.~(\ref{ineq3}) might actually hold \cite{Green}.

The tension between the two types of experiments might arise from underestimated systematic errors, but it may also be an actual sign of new physics. We focus on the latter case. Assuming that the discrepancy between the experimental results originates from an incomplete understanding of the physics behind neutron decay, the results of the two types of experiments can be reconciled if
\bea
{{\rm Br}(n\rightarrow p + {\rm anything})} \approx 99\% \ ,
\eea
while the remaining 1\% arises from \emph{\bf \emph{neutron dark decays}}, involving  at least one dark sector particle in the final state.

 \section{Neutron Dark Decay}
 
 To investigate how such decays could have gone unnoticed in other experiments, let us consider a general scenario of a neutron decaying to a final state $f$ with the sum of  final state particle masses equal to $M_f$.
 
Of course, for the neutron to undergo a dark decay,  $M_f$ has to be smaller than the neutron mass, i.e., $M_f < m_n$. The lower bound on $M_f$ is provided by experiments looking for neutron disappearance inside a nucleus. A neutron dark decay inside a nucleus $(Z, A)$ could  produce a daughter nucleus in an excited state $(Z, A\!-\!1)^*$, leading to its subsequent de-excitation with the emission of secondary particles, e.g.~gamma rays. A search for such signatures has been conducted by the SNO experiment \cite{Ahmed:2003sy} and the KamLAND experiment \cite{Araki:2005jt}, placing a constraint of $\tau_{n\to {\rm invisible}}> 5.8\times 10^{29}$ years, adopted by the PDG as the bound on the neutron invisible decay channel. 

However, if the condition $M_f > m_n - S_n$ is fulfilled, with $S_n$ being the neutron separation energy in a given nucleus, then the decay $(Z,A) \to (Z,A\!-\!1) + f$ is kinematically forbidden, while the neutron dark decay $n \to f$ is still allowed. Among all stable nuclei, the nucleus with the smallest neutron separation energy is $^9{\rm Be}$, with $S_n(^9{\rm Be}) = 1.664 \ {\rm MeV}$.  Thus, the requirement of $^9{\rm Be}$ stability enforces $M_f > m_n - 1.664 \ {\rm MeV}$, which leads  to the condition
\bea\label{con3}
937.900 \ {\rm MeV} <  M_f <  939.565 \ {\rm MeV} \ .
\eea
Since $937.9 \ {\rm MeV} > m_p - m_e$, the requirement in Eq.\,(\ref{con3}) also assures that proton would not undergo a dark decay. 

This opens the way to a whole new class of possible neutron decay channels: 
\bean
n \to \chi\,\gamma \ , \ \ \ \ n \to \chi\,\phi \ , \ \ \ \ n \to \chi\,e^+e^- \ ,  \ \ \ .\,.\,. \ \ \  ,
\eean
where $\chi$ is a dark fermion, $\phi$ is a dark scalar or a dark vector, and the ellipsis denotes other final states involving additional dark particles, photons and neutrinos. We now analyze the first two cases in more detail.

\subsection{${\boldsymbol {\rm Neutron \to dark \ particle +photon}}$}
This simplest case involves only one dark fermion $\chi$ and a monochromatic photon in the final state. The allowed range of masses for $\chi$, governed by Eq.~(\ref{con3}), is
 \bea\label{range3}
 937.900 \ {\rm MeV} <  m_\chi <  939.565 \ {\rm MeV} \, .
 \eea
 The energy of the corresponding monochromatic photon falls therefore within the range
 \bea\label{phE}
0 < E_\gamma < 1.664 \ {\rm MeV} \, . 
\eea
In the limit  $m_\chi \to m_n$, the photon energy $E_\gamma \to 0$. 

The dark fermion $\chi$ could be a dark matter particle, in which case its stability would require $m_\chi < m_p+m_e$, so that $\chi$ does not undergo beta decay through an off-shell neutron. In this dark matter case the allowed energy range for the photon reduces to $ 0.782 \ {\rm MeV} <E_\gamma < 1.664 \ {\rm MeV} $.

\noindent
 An effective Lagrangian for the decay $n \to \chi\,\gamma$ is 
 \bea\label{lageff113}
\mathcal{L}^{\rm eff}_1 \!\!\!&=&\!\!\! \bar{n}\,\big(i\slashed\partial-m_n +\tfrac{g_ne}{2 m_n}\sigma^{\,\mu\nu}F_{\mu\nu}\big) \,n\nonumber\\
&+&  \!\!\! \bar{\chi}\,\big(i\slashed\partial-m_\chi\big) \,\chi + \varepsilon \left(\bar{n}\,\chi + \bar{\chi}\,n\right) \ ,
\eea
where $g_n$ is the $g$-factor of the neutron and $\varepsilon$  is a model-dependent parameter with mass dimension one that governs the mixing between $\chi$ and $n$. The Lagrangian in Eq.\,(\ref{lageff113}) gives a neutron dark decay rate of
\bea\label{eff1}
\Delta\Gamma_{n\rightarrow \chi\gamma} = \frac{g_n^2e^2}{8\pi}\left(1-\frac{m_\chi^2}{m_n^2}\right)^3  \frac{m_n\,\varepsilon^2}{(m_n-m_\chi)^2} \  .
\eea
To explain the discrepancy between bottle and beam neutron lifetime experiments, $\Delta\Gamma_{n\rightarrow \chi\gamma} \approx \Gamma_n/100$, where 
$\Gamma_n$ is the total neutron decay rate in the Standard Model. 
 A phenomenologically viable particle physics model for the case $n \to \chi\,\gamma$ is discussed in Sec.~\ref{mod1} (Model 1).
\vspace{1mm}

\subsection{${\boldsymbol {{\rm Neutron \to two \ dark \ particles} }}$ }
A neutron dark decay with the final state consisting of only dark particles is realized by $n \to \tilde\chi^* \to \chi\,\phi$, where $\chi$ and $\tilde\chi$ are dark fermions and $\phi$ is a dark scalar ($\phi$ could also be a dark vector).  In this case the requirement in Eq.\,(\ref{con3})  takes the form
\bea
937.900 \ {\rm MeV} < m_\chi + m_\phi <  939.565 \ {\rm MeV} \ .
\eea
Since this condition involves only the sum of the $\chi$ and $\phi$ masses, $m_\chi$ does not need to be close to $m_n$, e.g. a scenario where $m_\chi \approx m_\phi \approx m_n/2$ is  allowed. However, nuclear stability requires that the mass of the intermediate $\tilde\chi$ satisfy
\bea
m_{\tilde\chi} > 937.9 \ {\rm MeV} 
\eea
to prevent $^9{\rm Be} \to \!\,^8{\rm Be} + \tilde\chi$. If, in addition, $|m_\chi - m_\phi| < m_p + m_e$, then both 
$\chi$ and $\phi$ cannot undergo beta decays.
\vspace{1mm}

\noindent
An effective Lagrangian describing $n\to \chi\,\phi$ is 
\bea\label{efflag2}
\mathcal{L}^{\rm eff}_{2}  \!\!\!&=& \!\!\!  \mathcal{L}^{\rm eff}_{1}(\chi \rightarrow \tilde\chi) +  \big(\lambda_\phi \,\bar{\tilde{\chi}}\, \chi\, \phi + {\rm h.c.}\big)\nonumber\\
&+&\!\!\!  \bar{\chi}\,\big(i\slashed\partial-m_{\chi}\big) \,\chi + \partial_\mu \phi^* \partial^\mu \phi - m_\phi^2\, |\phi|^2  \ , \ \ \ 
\eea
resulting in the neutron dark decay rate
\bea\label{rateb3}
\Delta\Gamma_{n\rightarrow \chi\phi} = \frac{|\lambda_\phi|^2}{16\pi}\sqrt{f(x, y)}\, \frac{m_n\,\varepsilon^2}{(m_n-m_{{\tilde\chi}})^2} \ ,
\eea
where  $f(x, y) =[(1-x)^2-y^2] \, [(1+x)^2-y^2]^3$,  $x=m_\chi/m_n$ and $y=m_\phi/m_n$. 

For $m_{\tilde\chi} > m_n$ the only available neutron dark decay channel is $n\to \chi\,\phi$ and $ \Delta\Gamma_{n\rightarrow \chi\phi} \approx \Gamma_n/100$ is needed to explain the neutron lifetime discrepancy. In the case $m_{\tilde\chi} < m_n$, the decay channel $n\to \tilde\chi\,\gamma$ is also allowed. The ratio of the corresponding dark decay rates is 
\bea\label{rate333}
\frac{\Delta\Gamma_{n\rightarrow \tilde\chi\gamma}}{\Delta\Gamma_{n\rightarrow \chi\phi}} = \frac{2g_n^2e^2}{|\lambda_\phi|^2} \frac{(1-\tilde{x}^2)^3 }{\sqrt{f(x, y)}}\ ,
\eea
where $\tilde{x} = m_{\tilde\chi}/m_n$. To account for the experimental discrepancy,
$ \Delta\Gamma_{n\rightarrow \chi\phi} + \Delta\Gamma_{n\rightarrow \tilde\chi\gamma} \approx \Gamma_n/100$.
A viable  model for the decay $n\to \chi\,\phi$ is provided in Sec.~\ref{mod2} (Model 2).

 \section{Particle Physics Models}\label{sec3}
 We emphasize that our  neutron dark decay proposal is very general and the models presented  below serve only as an illustration of the simplest scenarios. Theories with a more complex dark sector remain to be explored and, as discussed in Sec.~\ref{four3}, already their minimal realizations can solve several outstanding problems in  astrophysics.

 \subsection{Model 1 (${\boldsymbol{n\to \chi\,\gamma}}$)}\label{mod1}

\begin{figure}[h!]
\centering
\includegraphics[height=1.2in]{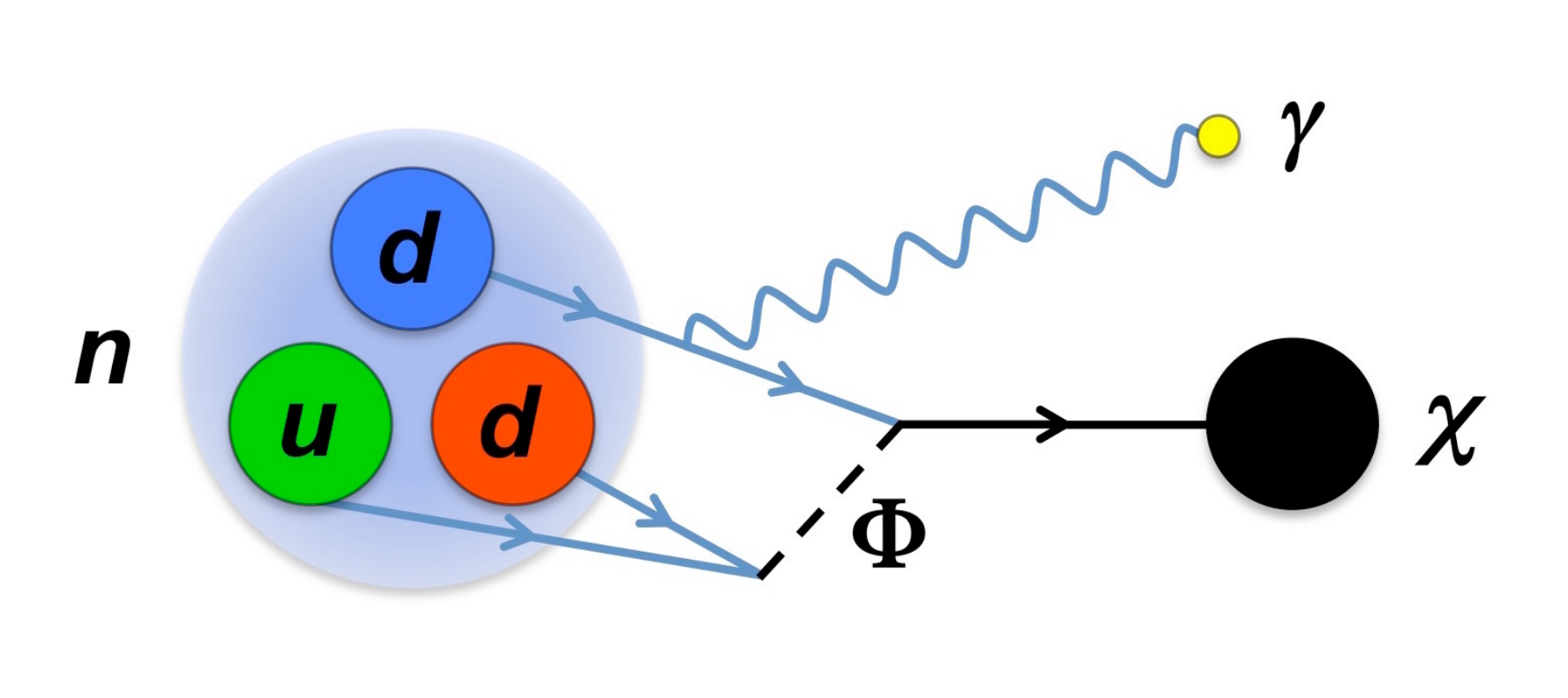}
\vspace{-2mm}
\caption{\small{Neutron dark decay $n \to \chi \, \gamma$ in Model 1.}} \vspace{1mm}
\label{fig:1}
\end{figure}

 The minimal model for the neutron dark decay requires only two new particles: a Standard Model singlet Dirac fermion $\chi$ and a scalar $\Phi$, chosen to be an ${\rm SU}(3)_c$ triplet, ${\rm SU}(2)_L$ doublet  and carrying  hypercharge $Y = -1/3$. 
 The Lagrangian of such a model is
 \bea\label{L13}
\mathcal{L}_{1} &\!\!\!=\!\!\!&   \Big[  \lambda_q \,\epsilon^{ijk}\, \overline{u^c_L}_{i}\, d_{Rj} \Phi_k + \lambda_\chi\Phi^{*i}\bar\chi \,d_{Ri}  + {\rm h.c.}\Big] \nonumber\\
&\!\!\!-\!\!\!& M_\Phi^2 \hspace{0.2mm}|\Phi|^2 - m_\chi \,\bar\chi\,\chi  \ , \ \ \ \ \ \ \ \ 
\eea
where $u^c_L$ is the charge conjugate of $u_R$. Assigning $B_\chi = 1$ and $B_\Phi=-2/3$, the theory conserves baryon number.
A diagram for $n\to \chi\,\gamma$ in this model is presented in Fig.~\ref{fig:1}.\\
\vspace{2mm}

\vspace{2mm}
The neutron dark decay rate is obtained by matching the Lagrangian in Eq.\,(\ref{lageff113}) with that in Eq.\,(\ref{L13}).  The result is given by Eq.\,(\ref{eff1}) with $\varepsilon = {\beta\,\lambda_q\lambda_\chi }/{M_{\Phi}^2}$, where $\beta$ is defined through $\langle 0| \,\epsilon^{ijk} (\overline{u^c_L}_{i} d_{Rj}) \,d_{Rk}^\rho |n\rangle = \beta \, \left({1+\gamma_5}\right)^{\rho}_{\, \sigma}  u^\sigma/2 $, with $u$ being the neutron spinor. Lattice calculations give $ \beta \approx 0.014 \ {\rm GeV}^3$ \cite{Aoki:2017puj}.\vspace{1mm}

There is a large parameter space available for which $\Delta\Gamma_{n\rightarrow \chi\gamma} \approx \Gamma_n/100$. For example, if one takes the mass of $\chi$ to be at  the lower end of the allowed range specified in Eq.\,(\ref{range3}), i.e., $m_\chi = 937.9 \ {\rm MeV}$, then the mass of $\Phi$ and the couplings in the model need to satisfy the relation 
\bea\label{cco3}
\frac{M_\Phi}{\sqrt{|\lambda_q\lambda_\chi|}} \approx 400 \ {\rm TeV} \ . 
\eea
Therefore, $\Phi$ easily avoids all collider bounds provided that ${M_\Phi} \!\!\gtrsim \!\!1 \ {\rm TeV}$. In addition, since $\chi$ is a Dirac fermion, it escapes the stringent constraints arising from neutron-antineutron oscillation \cite{Abe:2011ky} and dinucleon decay \cite{Gustafson:2015qyo} searches.

\subsection{Model 2 (${\boldsymbol{n\to \chi\,\phi}}$)}\label{mod2}

\begin{figure}[h!]
\centering
\includegraphics[height=1.3in]{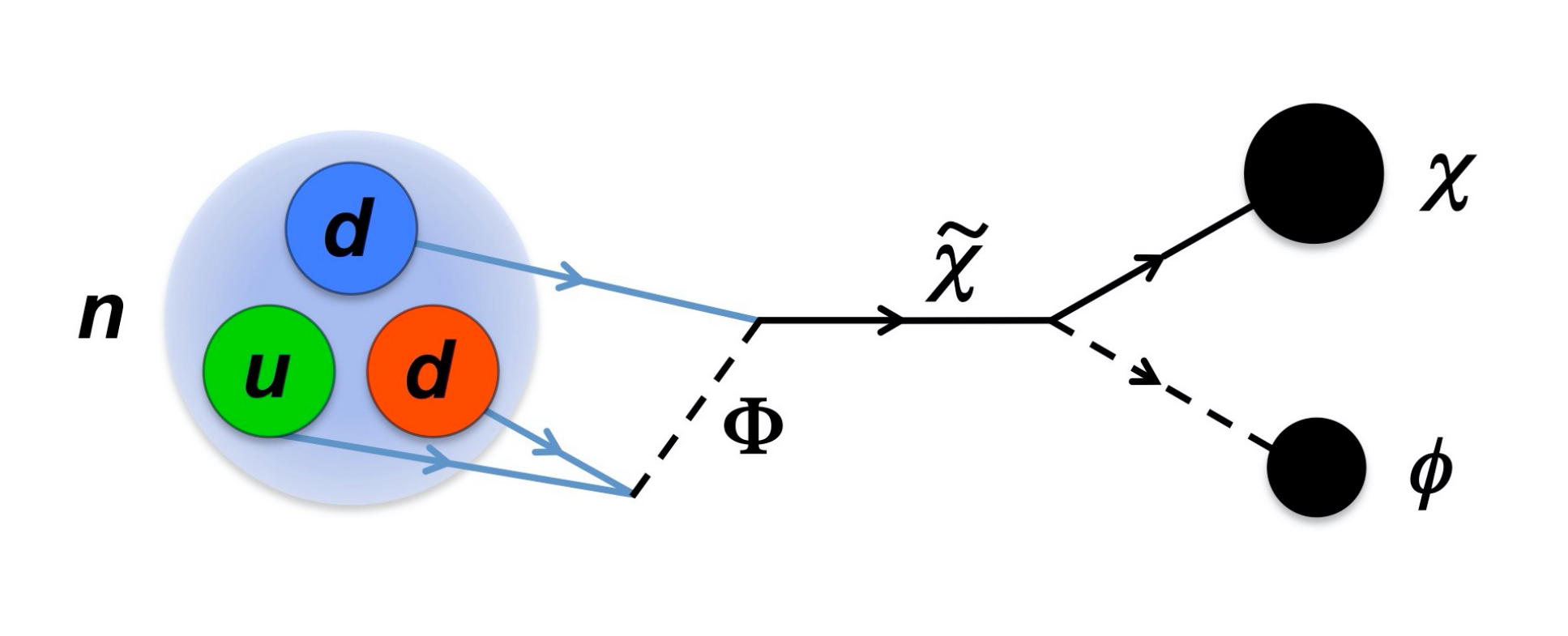}
\vspace{-8mm}
\caption{\small{Neutron dark decay $n \to \chi \, \phi$ in Model 2.}} \vspace{1mm}
\label{fig:2}
\end{figure}

The entirely dark decay of the neutron, involving two dark particles in the final state, requires adding four fields to the Standard Model:
the Dirac fermions $\chi$ and $\tilde\chi$, a scalar $\phi$ and the colored heavy scalar $\Phi$ introduced in the previous case.  The Lagrangian of the model resembles the one for Model 1 with $\chi$ substituted by $\tilde\chi$ and an additional interaction term between $\tilde\chi$, ${\chi}$ and $\phi$, i.e., 
\bea\label{333}
\mathcal{L}_{2}  &\!\!\!=\!\!\!& \mathcal{L}_{1}(\chi \rightarrow \tilde\chi)   +  ( \lambda_\phi  \,\bar{\tilde\chi}\, \chi \,\phi  + {\rm h.c.})  \nonumber\\
&\!\!\!-\!\!\!&  m_\phi^2\, |\phi|^2  -  m_\chi \,\bar\chi\,\chi   \ . 
\eea
Baryon number is conserved upon assigning $B_{\tilde\chi} = B_\phi=1$ and $B_\chi \!\!=\!\! 0$. 
The diagram for the neutron dark decay $n \to \chi \, \phi$ in this model  is shown schematically in Fig.~\ref{fig:2}.\\

After  matching the Lagrangians in Eqs. (\ref{efflag2}) and (\ref{333}), the rate for the  neutron dark decay $n \to \chi \, \phi$ is given by Eq.~(\ref{rateb3}) with $\varepsilon = {\beta\,\lambda_q\lambda_{\tilde\chi}}/{M_{\Phi}^2}$. The condition $\Delta\Gamma_{n\rightarrow \chi\phi} \approx \Gamma_n/100$, required to explain the neutron lifetime discrepancy when $m_{\tilde\chi} > m_n$, is satisfied for a wide range of parameters. In particular, adopting $m_\chi  = 937.9 \ {\rm MeV}$, $m_\phi \approx 0$ and $m_{\tilde\chi} =2 m_n$, the mass of $\Phi$ and the couplings of the model have to satisfy
\bea\label{cco23}
\frac{M_\Phi}{\sqrt{|\lambda_q\lambda_{\tilde\chi} \lambda_\phi|}} \approx 300 \ {\rm TeV} \ ,
\eea
again consistent with collider, neutron-antineutron oscillation and dinucleon decay constraints.

In the case $m_{\tilde\chi}<m_n$, the additional neutron decay channel $n \to \tilde\chi \,\gamma$ is also available and the sum of the two rates should add up to $\approx\Gamma_n/100$, with their ratio governed by Eq.\,(\ref{rate333}).

 \section{Theoretical Developments}\label{four3}
 
 Our work inspired several theoretical efforts to explore further  implications of neutron dark decays. This involved studying the physics of neutron stars in the presence of the new neutron decay  channels, constructing neutron dark decay models with a more complex dark sector including self-interactions, building models with dark decays of mesons, as well as inventing alternative, although related, ways to explain the neutron lifetime discrepancy. We discuss those theoretical ideas   below.

\subsection{Neutron star constraints}
The impact of neutron dark decays on neutron stars was considered in Refs.\,\cite{McKeen:2018xwc,Baym:2018ljz,Motta:2018rxp}. The resulting production of dark particles changes the energy density and pressure inside a neutron star, modifying its equation of state. This in turn changes the predictions for  the maximum allowed neutron star masses, since they are derived from integrating the Tolman-Oppenheimer-Volkoff equation that explicitly depends on the equation of state. 

It was shown that the observed neutron star masses ($2M_\odot$ for the heaviest neutron stars discovered) are allowed if strong repulsive self-interactions are present in the dark sector of our models. Such interactions are easily introduced in the representative Models 1 and 2 discussed in Sec.~\ref{sec3} by simply adding a dark vector boson coupled strongly to the dark particle $\chi$. 

Interestingly, a strongly self-interacting dark sector lies along the lines of the self-interacting dark matter paradigm, which was introduced two decades ago \cite{Spergel:1999mh} to solve the core-cusp and missing satellite problem of the $\Lambda {\rm CDM}$ model.

\subsection{Models with a self-interacting dark sector}

A model of this type was constructed in Ref.\,\cite{Cline:2018ami}, where a neutron dark decay involving a dark fermion and a dark photon in the final state was considered, i.e., $n \to \chi\, A'$. The effective Lagrangian  is
\bea\label{neweff3}
\mathcal{L}^{\rm eff} \!\!\!&=&\!\!\! \bar{n}\,\big(i\slashed{D}-m_n +\tfrac{g_ne}{2 m_n}\sigma^{\,\mu\nu}F_{\mu\nu}\big) \,n\nonumber\\
&+&  \!\!\! \bar{\chi}\,\big(i\slashed{D}-m_\chi\big) \,\chi + \varepsilon \left(\bar{n}\,\chi + \bar{\chi}\,n\right) \nonumber\\[2pt]
&-&\!\!\!\!\tfrac14{F'}_{\!\!\!\mu\nu} {F'}^{\mu\nu} - \tfrac{\delta}{2}{F}_{\!\mu\nu} {F'}^{\mu\nu} - \tfrac{1}{2}m_{A'}^2 {A'}_{\!\!\mu}{A'}^\mu \ ,
\eea
where the covariant derivative $D_\mu = \partial_\mu - i\,g'{A'}_{\!\!\mu}$. 
 It was shown that the strength of the dark photon coupling to the dark particle $\chi$, governed by the parameter $g'$ and resulting in repulsive interactions between the $\chi$ particles, can be chosen such that the neutron lifetime discrepancy is explained and, at the same time, all astrophysical bounds are satisfied, including constraints from neutron stars, galaxy clusters, cosmic microwave background, Big Bang nucleosynthesis and supernovae. If the dark particle $\chi$ in this model is stable, it can contribute to the dark matter in the universe, but cannot account for all of the dark matter.

Many of the astrophysical  constraints are alleviated if one assumes non-thermal dark matter production. This was shown in Ref.\,\cite{Karananas:2018goc}, where a model for the neutron dark decay $\,n \to \chi \, \phi\,$ was constructed, based on our Model 2, but with a dark boson introduced to mediate large self-interactions of $\chi$. The Lagrangian for the dark sector is
\bea
\mathcal{L}_D&\!\!\!=\!\!\!&  g\,\bar{\chi}\,\slashed{\hspace{0mm}Z}_D \,\chi + ( \lambda_\phi  \,\bar{\tilde\chi}\, \chi \,\phi  + {\rm h.c.}) \nonumber\\
&\!\!\!-\!\!\!& i\, g\,Z_D^{\,\mu} \,\big(\phi^*\partial_\mu \,\phi - \phi\,\partial_\mu \phi^* \big) \ .
\eea
There exists a choice of parameters  for which this model
 satisfies neutron star constraints, remains consistent with all other astrophysical bounds and $\chi$ makes up all of the dark matter in the universe. In addition, due to the self-interactions of $\chi$, the model is shown to solve the small-scale structure problems of the $\Lambda {\rm CDM}$ model.

\subsection{Hadron dark decays} 
The idea of dark decays can be applied also to other neutral hadrons. In Ref.\,\cite{Barducci:2018rlx} it was argued that the mesons $K_L^0$ and $B^0$ can decay to dark sector particles at measurable rates. An explicit model was constructed with a dark sector consisting of several families  of dark fermions.  An analogous mechanism that prevents neutron beta decays in neutron stars, i.e., Pauli blocking, also forbids neutron dark decays inside a neutron star in this model.

\subsection{Baryogenesis} 
It has recently been shown that the model addressing the neutron lifetime puzzle based on the Lagrangian in Eq.\,(\ref{neweff3}) provides a successful framework for low-scale baryogenesis \cite{Bringmann:2018sbs}.
In addition, a  model very similar to our Model 2, with couplings of $\tilde\chi$ to other quark flavors and a Majorana (instead of Dirac) fermion $\chi$, has been proposed  in the context of low-scale baryogenesis as well \cite{Elor:2018twp}.

\subsection{Related solutions}
Taking into consideration  only the experimental data for $g_A$ from experiments performed after the year 2002, the bottle neutron lifetime is favored \cite{Czarnecki:2018okw}. Based on this observation, explanations of the neutron lifetime discrepancy have been put forward in which it is the bottle lifetime that is equivalent to the Standard Model prediction for $\tau_n$. The difference in outcomes of the bottle and beam measurements is explained via neutron-mirror neutron oscillations resonantly enhanced in large magnetic fields thus affecting  only beam measurements \cite{Berezhiani:2018eds}, or by invoking a sizable Fierz interference term canceling the dark decay contribution to the neutron decay rate \cite{Ivanov:2018vit}.

\section{Experimental Searches}

Several experimental efforts have been undertaken directly after our results were announced, searching specifically for the signatures we proposed.

\subsection{${\boldsymbol {\rm Neutron \to dark \ matter +photon}}$}
Within the first few weeks after our results became public, a dedicated experiment was performed at the Los Alamos UCN facility looking  for the monochromatic photon in the neutron dark decay $n \to \chi\,\gamma$ \cite{Tang:2018eln}. The search was sensitive to final state photons with energies $0.782 \ {\rm MeV} < E_\gamma < 1.664 \ {\rm MeV}$ and challenged the case ${\rm Br}(n\to \chi\,\gamma) \approx 1 \%$ at a significance level of $2.2 \, \sigma$. 
The remaining photon energy range, i.e., $E_\gamma < 0.782 \ {\rm MeV}$, is left to be explored.

\subsection{${\boldsymbol {{\rm Neutron \to dark \ particle} +{e^+e^-}}}$} 
Another dedicated experiment, also performed at the Los Alamos UCN facility, looked  for $e^+e^-$ pairs from the neutron dark decay $n\to \chi\,e^+e^-$ \cite{Sun:2018yaw}. This search excluded the case with ${\rm Br}(n\to \chi\,e^+e^-) \approx 1 \%$ for the electron-positron energy range $E_{e^+e^-} \gtrsim 2\,m_e + 100 \ {\rm keV}$ with a confidence of nearly $100\%$. The remaining $100 \ {\rm keV}$ energy window was beyond experimental  sensitivity.

\subsection{\bf Nuclear dark decays} 
There exists a number of unstable nuclei for which the neutron separation energy is smaller than for $^9{\rm Be}$, i.e., $S_n < 1.664 \ {\rm MeV}$. Those include $^7{\rm H}$, $^{11}{\rm Li}$, $^{11}{\rm Be}$, $^{13}{\rm Li}$, $^{14}{\rm B}$, $^{15}{\rm C}$, $^{16}{\rm Be}$, $^{17}{\rm B}$, $^{17}{\rm C}$, $^{19}{\rm B}$, $^{19}{\rm C}$, $^{22}{\rm C}$, $^{22}{\rm N}$, as well as heavier ones. For these particular nuclei a neutron dark decay can lead to nuclear dark decays if the final state dark particle mass $m_\chi$ falls within the range
\bea
937.9 \ {\rm MeV} < m_\chi < m_n - S_n \ .
\eea
We proposed to seach for such nuclear dark decays in our original paper \cite{Fornal:2018eol}, focusing on the corresponding signatures for $^{11}{\rm Li}$, for which $S_n(^{11}{\rm Li}) = 0.396\ {\rm MeV}$. In that case the decay chain $^{11}{\rm Li} \to \!\,^{10}{\rm Li} +\chi \to \!\,^{9\,}{\rm Li} +n+\chi $ is allowed and the $^9{\rm Li}$ long lifetime could be used to discriminate against the background from $^{11}{\rm Li}$ beta decays. However, $^9{\rm Li}$ can be produced also in beta-delayed deuteron emission \cite{KELLEY201288,Raabe:2008rj} and the distinction between this and the dark channel would be extremely difficult.

It was argued in Ref.\,\cite{Pfutzner:2018ieu} that, from an experimental point of view, there is a much better candidate: $^{11}{\rm Be}$, for  which  $S_n(^{11}{\rm Be}) = 0.502 \ {\rm MeV}$. It was also suggested that the presence of an unexpectedly high number of 
 $^{10}{\rm Be}$ in $^{11}{\rm Be}$ decays described in Ref.\,\cite{Riisager:2014gia} might in fact be a sign of the neutron dark decay $n \to \chi\,\phi$ like in our Model 2, leading to the nuclear dark decay 
 \bea\label{cons34}
^{11}{\rm Be}\to \!\,^{10}{\rm Be} + \tilde\chi^*  \to  \!\,^{10}{\rm Be} + \chi + \phi\ ,
\eea
and not necessarily,  as initially conjectured, due to an enhanced $\beta p$ channel resulting from an unknown resonance. 

In addition, it was shown in Ref.\,\cite{Ejiri:2018dun} that the nuclear dark decay in Eq.\,(\ref{cons34}) is consistent with the observed Standard Model decay rates of $\,^{11}{\rm Be}$ as long as  $m_{\tilde\chi} > m_n - S_n(^{11}{\rm Be})$ , i.e.,
\bea\label{Eji}
m_{\tilde\chi} > 939.064 \ {\rm MeV} \ . 
\eea
This condition is obviously satisfied in the model with a self-interacting dark sector of Ref.\,\cite{Karananas:2018goc}, where the $\tilde\chi$ mass was chosen  to be $m_{\tilde\chi} = 800 \ {\rm GeV}$.

Very recently, an experiment at the CERN-ISOLDE laboratory  was performed \cite{Pfutzner:2018ieu,ISOLDE} with the goal of determining 
whether the final state of $^{11}{\rm Be}$ decays contains protons in the final state or not. The results have not yet been published.

\subsection{Ongoing beam measurements}

There are currently two operating beam experiments measuring the neutron lifetime, the first one at the National Institute of Standards and Technology (NIST) \cite{NIST2009,NIST} and the second one at the Japan Proton Accelerator
Research Complex (J-PARC) \cite{Nagakura:2017xmv,Japan}. If those experiments provide results consistent with the current beam average, the tension between bottle and beam measurements will increase, supporting the viability of models presented here.

\subsection{Expanding the scope of  bottle experiments}
Perhaps the most straightforward, although technically challenging way to tackle the neutron lifetime puzzle  would be to modify the existing bottle experimental  setup.
Including a proton detection system in bottle experiments would enable  measuring the branching fraction ${\rm Br}(n\to p + {\rm anything})$ independently of the beam experiment.
Such modification would enable a direct test of the premise that the difference of outcomes between the bottle and beam measurements is due to neutron decays that do not produce a proton, without any dependence on the specific model realization of the non-proton final state.

\section{Final Remarks}

Given the theoretical and experimental developments related to our proposal, Model 2 with a self-interacting dark sector seems like a very promising candidate theory for explaining the neutron lifetime discrepancy. This model is not only consistent with all current experimental constraints, but it is also interesting from a theoretical perspective, with its solution to the small-scale structure problem and perhaps a novel mechanism for baryogenesis.

Even if the neutron lifetime puzzle gets resolved by future higher precision bottle and beam measurements, dark decays of the neutron at a smaller rate will still be allowed and certainly interesting to consider. 
It would be incredible if the good old neutron became the key to unraveling the mystery of the dark  side of our universe.

\section*{Acknowledgements}

This research was supported in part by the DOE Grant No.~${\rm DE}$-${\rm SC0009919}$.

\end{document}